\documentstyle[aps,graphicx,tighten,preprint]{revtex}

\begin{document}
\title{Proposal to test quantum mechanics against
 macroscopic realism using continuous variable entanglement: a definitive
  signature of a  
Schr\"odinger cat\\ }
 \vskip 1 truecm
\author{M. D. Reid\\ }
\address{Physics Department, University of Queensland, Brisbane, Australia\\ }  	  
\date{\today}
\maketitle
\vskip 1 truecm
\begin{abstract}
In the Schr\"odinger-cat gedanken experiment a cat 
is in a quantum superposition of two macroscopically distinct states, 
alive and dead.
The paradoxical    
interpretation of quantum mechanics is  
that the cat is not in one state or the other, 
alive or dead, immediately prior to its measurement. 
Because of this  
apparent defiance of macroscopic reality, 
quantum superpositions of states macroscopically distinct have  
generated  
much interest. 
Here we address the issue of proving a contradiction with   
macroscopic reality objectively, 
through the testable predictions of quantum 
mechanics. We consider 
the premises of macroscopic reality 
(that the ``cat'' is either ``dead'' 
or ``alive'', the measurement indicating which) and macroscopic locality 
(that simultaneous measurements some distance away cannot 
induce the macroscopic change, ``dead'' to ``alive'' and vice versa,  
to the ``cat''). 
The predictions of quantum mechanics for certain 
states, generated using states exhibiting continuous variable 
entanglement, are shown to be   
incompatible with the predictions of all theories based on 
this dual premise. Our proof is along the lines of 
Bell's theorem, but  
where all relevant measurements give macroscopically distinct results.   

\end{abstract}
\narrowtext
\vskip 0.5 truecm

Schr\"odinger$^{\cite{1}}$ raised the issue of the existence and 
 interpretation of the quantum superposition of two macroscopically 
 distinct states in his famous 
Schrodinger-cat gedanken experiment. A particle is in a 
quantum superposition of having escaped the nucleus, or 
 otherwise. The presence of the particle outside the nucleus will  
trigger a lethal device that will kill a cat located 
in a box. An observer later looks into the box to determine the 
state of the cat, whether dead or alive. 
The application of quantum mechanics, to all stages of 
the sequence of interactions, would ultimately predict
 the cat to 
be in a superposition of a state $|1\rangle$, where the cat is dead, and a 
state $|-1\rangle$, 
where the cat is alive. 

It is a basic feature of quantum mechanics that the  
quantum superposition state $(|+\rangle+|-\rangle)/\sqrt2$ 
cannot be regarded as classical 
mixture, where 
the system is considered to be in state $|+\rangle$ 
with probability $1/2$, or in state $|-\rangle$ with 
probability $1/2$. 
Yet to say in this case 
that the cat cannot be considered to be dead or alive prior to its 
measurement, here the  
observer opening the box to view the state of the cat, would seem 
nonsensical.

 The recent experimental evidence for the generation$^{\cite{2}}$ of 
a quantum superposition of two 
macroscopically distinct states
 makes it timely to consider definitive signatures of a Schr\"odinger 
 cat state. The macroscopic
  superposition state is of basic interest 
 because of its 
 paradoxical interpretation: 
 that a macroscopic object (a cat) was not actually  
 in one of two macroscopically distinct states 
 (dead or alive), prior to its measurement. 
 An important issue is then not simply the 
existence of the macroscopic superposition state, but its 
interpretation.
 So far evidence, presented within the framework of quantum mechanics, has 
 been for the existence of these states$^{\cite{2,3,4,5,6}}$. The fundamental 
 issue that there could be an 
 alternative theory or interpretation of quantum mechanics, in which 
 the ``cat'' is either ``alive'' or ``dead'' , the measurement 
 indicating which, is a fundamental one and also needs to be examined. 
 
 Combining the previous approaches of Bell$^{\cite{7}}$ and 
 Leggett and Garg$^{\cite{8}}$, 
 I show here that the 
 predictions of  
 quantum mechanics for certain quantum states are incompatible with 
 a large class of such alternative theories, those embodying the 
 combined premise of macroscopic realism and macroscopic locality as 
 defined below. In doing so I  
  show that certain quantum mechanical 
 Schr\"odinger cats can irrefutably defy  
 macroscopic reality, and so propose a definitive 
 Schr\"odinger's cat experiment in which the macroscopic paradoxical 
 nature of the ``cat'' can be proven objectively.

We first must define
 what is meant  
 by macroscopic reality in the context of  
 Schr\"odinger's gedanken 
 experiment or experiments analogous to it. Consider a macroscopic 
 system (I will call the system the ``cat'') 
 giving one of two macroscopically distinct outcomes (``dead'' 
 or ``alive'') for that system upon 
 measurement. 
   First, following Leggett and Garg$^{\cite{8}}$,   
  the   
   premise of
   macroscopic reality is defined to imply the following:   
   that the macroscopic system (the cat)
   is actually   
in one of two macroscopically distinct states, dead or 
 alive, prior to its measurement, and that
  the measurement simply    
 elucidates 
 which of the two states, dead or alive, the cat was in. 
 We introduce a hidden 
 variable $\lambda$ to denote the predetermined state of the cat, 
 $\lambda=+1$ representing the state ``dead'' and $\lambda=-1$ 
 representing the state ``alive''.
 Second 
 it is postulated 
 that a measurement 
 performed simultaneously on a second macroscopic system (or  
 cat)
  cannot induce a macroscopic 
 change (dead to alive or vice versa) to the state $\lambda$, or to the  
 result of the measurement of the first cat. 
 Where the two cats are spatially separated, this last premise may 
 be termed ``macroscopic locality''$^{\cite{9}}$ . 

 It is the assumption of macroscopic reality    
 that the measurement merely elucidates,  
 and does not induce, the state of the 
 cat. I propose that this assumption naturally carries with it the 
 above premise of macroscopic 
 locality, that    
 measurements on other cats cannot instantaneously induce a 
 change to the state of the first cat.

The idea of testing quantum mechanics against all theories based on 
certain classical premises was put forward by Bell$^{\cite{7}}$ in 1966. 
Bell's result however applies to 
quantum 
superpositions of states that are only microscopically distinct. 
Leggett and Garg$^{\cite{8}}$ have since shown the incompatibility   
  of quantum mechanics with 
  a dual premise called macroscopic realism and macroscopic 
  noninvasiveness of 
  measurement.
While the confirmation of this quantum prediction would 
be significant, the result would not exclude the possibility 
that the cat is either 
dead or alive, 
provided one accepts that the measurement of a macroscopic system  
alters its subsequent evolution.

 I now discuss how  
 quantum mechanics can predict the existence of a macroscopic 
 superposition state 
 that defies the dual macroscopic reality-locality premise 
 above. Consider measurements made simultaneously on each of 
  two macroscopic 
systems, cat $A$ and cat $B$ (Figure 1). It  
is possible to perform, on each macroscopic system,   
only one of two measurements$^{\cite{10}}$, designated 
by $\theta$ and 
$\theta'$ for cat $A$, and by $\phi$ and $\phi'$ for cat $B$.
 To provide an analogy with the original gedanken experiment, we
  might picture that the observer must view the cat through   
 an optical filter, so we call these the 
``blue'' measurement (corresponding to $\theta$ and $\phi$)
 and the ``green'' measurement (corresponding to $\theta'$ and 
 $\phi'$).

For each measurement we have two possible outcomes, 
denoted by $+1$ and $-1$, and these   
  outcomes are 
macroscopically distinct, for both blue and green measurements, and for 
both cats $A$ and $B$. 
By this we mean 
that the two possible outcomes of the measurement 
(certain eigenvalues of the quantum 
measurement operator) 
correspond to macroscopically distinct states of the macroscopic 
system. This situation of macroscopically distinct outcomes is directly 
    parallel to that of the dead and alive results of measurement 
    of Schr\"odinger's cat. I now assume the premise above, that the cat 
    $A$ is either dead (result $1$) or 
    alive (result 
    $-1$),
     for the blue measurement,
     immediately prior to the measurement. I introduce 
    the hidden variable $\lambda_{blue}^{A}¥$ to represent the 
    predetermined nature of the cat, for the blue measurement. 
     Here $\lambda_{blue}^{A}¥=1$ 
    represents the cat dead and $\lambda_{blue}^{A}¥=-1$ 
    represents the cat alive. According to our premise
     the result of the measurement is given directly  
      by the value assumed by the hidden variable.

    Two different measurements, with macroscopically distinct 
    results, can be performed at $A$ and at $B$ so 
    that there are in total four hidden variables $\lambda_{blue}^{A}$, 
    $\lambda_{green}^{A}$,$\lambda_{blue}^{B}¥$ and $\lambda_{green}^{B}¥$ each 
    assuming a value either $1$ or $-1$. Substitution of all 
    possible values shows:
$    -2\leq \lambda_{blue}^{A}\lambda_{blue}^{B}-
     \lambda_{blue}^{A}\lambda_{green}^{B}
     +\lambda_{green}^{A}\lambda_{blue}^{B}
     +\lambda_{green}^{A}
     \lambda_{green}^{B}\leq 2.
$ The 
 prediction for the averages calculated over many 
 experimental runs follows directly. We introduce:    
 $E(B,B)=\langle\lambda_{blue}^{A}\lambda_{blue}^{B}\rangle$, the expectation 
 value for blue measurements at both $A$ and $B$;
  $E(B,G)=\langle\lambda_{blue}^{A}\lambda_{green}^{B}\rangle$, the expectation 
  value for blue 
  measurement at $A$ and a green measurement at $B$; and so on.   
     \begin{equation}
   E=E(B,B) -E(B,G)+E(G,B)
   +E(G,G)\leq 2
\end{equation}
This result is a Bell inequality, but since the outcomes of all 
relevant measurements are macroscopically distinct, the inequality is 
derivable from our premise of 
macroscopic realism-locality$^{\cite{11}}$.

 I present a quantum state 
violating this inequality for macroscopically distinct outcomes. 
We consider initially (Figure 1) the situation where each 
macroscopic system ($cat$)
is a macroscopic  
field of fixed frequency comprised of 
two orthogonal polarisation directions. 
Each quantised field mode of a given frequency and polarisation is 
 equivalent to a quantum harmonic 
oscillator system. We introduce 
boson operators
 ${\bf \hat{a}_{-}}$ and  ${\bf \hat{a}_{+}}$ for the 
 two orthogonally polarised modes 
 of cat $A$; similarly we have  
 ${\bf \hat{b}_{-}}$ 
and  ${\bf \hat{b}_{+}}$ for $B$.

  On each system, $A$ and $B$, a measurement is 
made with a two-channeled polariser which transmits light polarised at angle 
$\theta/2$ for $A$, and $\phi/2$ for $B$. 
The transmitted mode for $A$ is represented 
by  
$\hat{c}_{+}={\bf \hat a_{+}}\cos(\theta/2)+ {\bf \hat 
     a_{-}}\sin(\theta/2)$, and the orthogonally- 
    polarised field mode by    
$\hat{c}_{-}={\bf \hat  a_{+}}\sin(\theta/2)- {\bf \hat 
a_{-}}\cos(\theta/2)$. 
\begin{figure}
 \includegraphics[scale=.7]{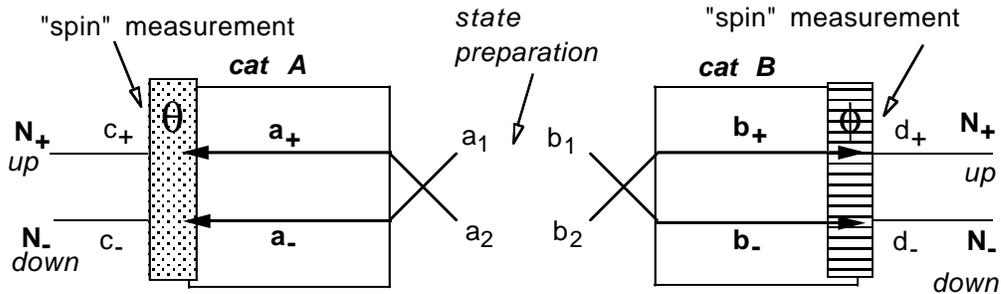}\\
\caption[diagrams]
{Arrangement to demonstrate that Schr\"odinger cats $A$ and $B$ defy 
a macroscopic reality. The number $N_{+}$ of particles (intensity) 
polarised ``up'' 
is either 
macroscopically larger (result $+1$), or macroscopically smaller 
(result $-1$),  
 than  $N_{+}$, the number of particles (intensity) polarised ``down''. 
}% 
\end{figure}    

The 
detection of the fields $\hat{c}_{\pm}$   
allows determination of the total number  
$N_{+}^{A}=\hat{c}^{\dagger}¥_{+}\hat{c}_{+}$, 
of particles in the transmitted $up$
 direction for $A$, and 
the total number of particles, 
$N_{-}^{A}¥=\hat{c}^{\dagger}_{-}\hat{c}_{-}$, in the orthogonal or 
$down$ 
direction for $A$.
Our measurement on cat $A$ is of the particle number 
difference 
\begin{equation}
2\hat{S}_{z}^{A}¥(\theta)=¥N_{+}^{A}-N_{-}^{A}
=\hat{c}^{\dagger}¥_{+}
\hat{c}_{+}-\hat{c}^{\dagger}_{-}\hat{c}_{-}
\end{equation}
where using the Schwinger representation we have a 
direct equivalence to spin measurements. For measurement on cat $B$ 
we 
have similar definitions  
$\hat{d}_{+}={\bf \hat b_{+}}\cos(\phi/2)+ {\bf \hat 
b_{-}}\sin(\phi/2)$ 
and $\hat{d}_{-}={\bf \hat b_{+}}\sin(\phi/2)- {\bf \hat 
b_{-}}\cos(\phi/2)$. The particle number difference for cat $B$ is 
$2\hat{S}_{z}^{B}¥(\phi)=N_{+}^{B}-N_{-}^{B}$
where $N_{+}^{B}=\hat{d}^{\dagger}¥_{+}
\hat{d}_{+}$ and $N_{-}^{B}=\hat{d}^{\dagger}_{-}\hat{d}_{-}
$.

It is necessary that the fields, our ``cats'', be  
macroscopic states, of large 
particle or photon number. This experiment,
 with a macroscopic number of particles 
incident on the polarisers, is then a macroscopic 
version of the original Bell inequality experiment$^{\cite{7}}$, the 
original Bell proposal  
involving only one particle incident on each polariser (or 
spin analyser). 
While such macroscopic Bell experiments have 
been examined previously$^{\cite{12,13,14,15}}$ by Mermin, Peres  
and others, it is an 
crucial requirement of 
our experiment that the relevant outcomes for both blue ($\theta,\phi$) and green 
($\theta',\phi'$) measurements, on both cats, are macroscopically distinct.
 
 As with Schrodinger's original gedanken experiment, we propose to 
 generate (Figure 1) the 
 macroscopic state, for ${\bf \hat a_{\pm}}¥$ and 
 ${\bf \hat b_{\pm}}¥$, from a microscopic quantum state  
$|\psi\rangle$ for two field modes
 $\hat{a}_{1}¥$ and $\hat{b}_{1}¥$. We  
require that this original state be entangled, 
and as one example we consider the 
circular superposition of coherent states$^{\cite{16}}$  
    \begin{equation}
        |\psi\rangle = N \int_{0}^{2\pi}
        |r_{0}e^{i\varsigma}\rangle_{a_{1}¥}
        |r_{0}e^{-i\varsigma}\rangle_{b_{1}¥} d\varsigma 
\end{equation}
where $r_{0}=1.1$, $N^{2}¥=e^{2r_{0}^{2}}/4\pi^{2}I_{0}(2r_{0}^{2})$
 and 
$|{\gamma}\rangle_{k}$ is the coherent state for mode $k$.
 We introduce a second pair of macroscopic quantum fields  
 $\hat a_{2}$ and $\hat b_{2}$, in coherent states 
$|\alpha\rangle_{a_{2}¥}$ and  $|\beta\rangle_{b_{2}¥}$ respectively, where 
$\alpha$, $\beta$ are real and large. 
Fields  $\hat a_{1}, \hat a_{2}¥$ are combined (using beam splitters 
or polarisers) 
to give macroscopic fields
${\bf \hat a_{-}}=(\hat a_{1}-\hat a_{2})/\sqrt{2}$ and  
    ${\bf \hat a_{+}}=i(\hat a_{1}+\hat a_{2})/\sqrt{2}
$  incident on 
    the polariser 
    for $A$. Similarly $\hat b_{1}, \hat b_{2}$ are mixed 
    to give macroscopic outputs ${\bf \hat b_{\pm}}¥$ for $B$.
         The  
    systems being measured, the 
    fields ${\bf \hat a_{\pm}},{\bf \hat b_{\pm}}$ incident on the 
    polarisers, are macroscopic.
     The total size of the system $A$ is  
    $\hat{S}_{A}=N_{+}^{A}+N_{-}^{A}=\hat{a}^{\dagger}¥_{1}
\hat{a}_{1}+\hat{a}^{\dagger}_{2}\hat{a}_{2}$, which as $\alpha \rightarrow\infty$ 
    is dominated by the Poissonian probability distribution 
    mean $\alpha^{2}¥$. The fields ${\bf \hat a_{+}},{\bf \hat 
    a_{-}}$ are  
    individually 
    similarly macroscopic with  
    mean photon number $\alpha^{2}¥/2$. 

We choose to write this macroscopic state $|\Psi\rangle$ using as a basis 
(the Pointer basis) 
the eigenstates $|m\rangle,|n\rangle,|S_{A}\rangle,|S_{B}\rangle$ of the 
measured quantities $2\hat{S}_{z}^{A}¥(\theta)=N_{+}^{A}-N_{-}^{A}$, 
$2\hat{S}_{z}^{B}¥(\phi)=N_{+}^{B}-N_{-}^{B}$, 
$\hat{S}_{A}=N_{+}^{A}+N_{-}^{A}$, 
$\hat{S}_{B}=N_{+}^{B}+N_{-}^{B}$ respectively.
 This is done by noting 
$\hat{c}_{+}=i(\hat a_{1}\exp(-i\theta/2)+\hat 
a_{2}\exp(i\theta/2))/\sqrt 2$ and 
$\hat{c}_{-}=(-\hat a_{1}\exp(-i\theta/2)+\hat 
a_{2}\exp(i\theta/2))/\sqrt 2$. 
      \begin{eqnarray}
	|\Psi\rangle &=& 
	\sum_{m,n=-\infty}^{\infty}\sum_{S_{A},S_{B}}
	A(m,n,S_{A},S_{B}) 
 |m\rangle|n\rangle|S_{A}\rangle|S_{B}\rangle 
\end{eqnarray}
Values for $S_{A}, S_{B}$ 
must ensure positive $\hat{c}^{\dagger}¥_{\pm}
\hat{c}_{\pm}$,
$\hat{d}^{\dagger}¥_{\pm}
\hat{d}_{\pm}$. The joint probability $P(m,n)$ of outcome 
$m,n$, for measurements $N_{+}^{A}-N_{-}^{A}$, $N_{+}^{B}-N_{-}^{B}$, is   
$P(m,n)=\sum_{S_{A},S_{B}}|A(m,n,S_{A},S_{B})|^{2}$.
Results show complete symmetry between positive and negative $m$  
values: $A(-m,n,S_{A},S_{B})=A(m,n,S_{A},S_{B})$; also for $n$.
 The crucial point is that for any fixed but arbitrary, positive $N_{0}$, 
the 
amplitude $A(m,n,S_{A},S_{B})$ where $-N_{0}\leq m,n\leq N_{0}$ can be 
made arbitrarily small
 by increasing 
$\alpha,\beta$. In this way we obtain a superposition 
of states with
 $m<N_{0}¥$ and states with $m>N_{0}¥$. This is true 
for all choices of basis $\theta,\phi$, and 
as $N_{0}$ becomes large we have a multi-faceted (blue and green) 
Schrodinger-cat state$^{\cite{17}}$. Importantly the cat-state is 
prepared prior to the choice of measurement angles $\theta,\phi$.

    To give our proposed definitive proof of the Schrodinger cat we  
    demonstrate that in the asymptotic limit, $\alpha,\beta \rightarrow 
    \infty$, 
   we have two macroscopically distinct outcomes 
   for measurements $S_{z}^{A}¥(\theta)$ (and $S_{z}^{B}¥(\phi)$), 
   yet can still violate the Bell inequality (2).
   Consider three regions of 
    outcome: first, where the result $m$ 
    for the particle number difference 
    $N_{+}-N_{-}$ 
    satisfies  $m> N_{0}¥$, result designated $1$; second, 
    where    $m< -N_{0}¥$ (result $-1$); and third, 
    $-N_{0}\leq m \leq N_{0}¥$ (result $0$). 
    For $N_{0}¥$ macroscopic, the outcomes 
    $1$ and $-1$ are macroscopically distinct in particle number $m$. For 
    arbitrarily large, fixed 
    $N_{0}¥$, the 
    probability of result $0$, 
    $\sum_{m=-N_{0}}^{N_{0}}\sum_{n=-\infty}^{\infty}¥P(m,n)$, 
    becomes increasingly negligible 
    as $\alpha, \beta\rightarrow \infty$, 
    while a clear violation of our inequality (1) in this asymptotic limit 
   is maintained, at $E=2.03$$^{\cite{18}}$ (Figure 2).

Plots of the probability distribution $P(m,n)$ do indeed reveal the 
asymptotic macroscopic limit where the shape of the distribution is unchanged. 
This final asymptotic shape $P_{a}(x,y)$ is dependent only on  
$x=m/\alpha,y=n/\beta$. 
As 
$\alpha=\beta$ increase, this 
probability function moves outward along the photon number $m$ and $n$ 
axes.

The macroscopic asymptotic limit can be checked analytically by 
 expanding the operators for the coherent fields$^{\cite{9}}$ as 
 $\hat a_{2}=\alpha+\delta \hat a_{2}$. In the large 
 $\alpha,\beta$ limit leading terms are
$2\hat{S}_{z}^{A}¥(\theta)
\sim\alpha \hat X_{\theta}^A
$ (similarly
$2\hat{S}_{z}^{B}(\phi)\sim\beta \hat X_{\phi}^{B}$).  
 The quadrature phase amplitudes 
 $\hat X_{\theta}^A=\hat a_{1}\exp(-i\theta)+\hat a_1^\dagger \exp(i\theta)$ 
and $\hat X_{\phi}^B=\hat b_1\exp(-i\phi)+\hat b_1^\dagger \exp(i\phi)$ 
    are linear combinations of the ``position'' and ``momentum''
     variables of the    
    harmonic oscillators $\hat{a}_{1}$, $\hat{b}_{1}$, respectively. 
    Our measurement 
    \begin{figure}
\includegraphics[scale=.7]{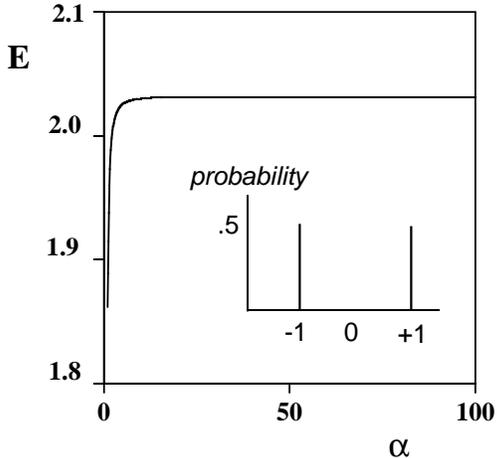}\\
\caption[diagrams]
{Quantum prediction of $E$ versus $\alpha=\beta$, with 
$\theta=0,\phi=-\pi/4,\theta'=\pi/2,\phi'=-3\pi/4$.
 The inset shows the probabilities, for 
all choices of $\theta$, and $\phi$, of obtaining 
results $-1$,$0$, and $+1$, at $A$ or $B$, in the large $\alpha,\beta$ limit.
}% 
\end{figure}

    $2\hat{S}_{z}^{A}¥(\theta)$ in Figure 
 1 for large $\alpha$ is operationally equivalent to the balanced 
homodyne detection measurement$^{\cite{19}}$ of 
$\hat X_{\theta}^{A}$, 
except that here all fields are macroscopic prior to 
the selection of the measurement angle $\theta$. 
 The asymptotic form, $P_{a}¥(x,y)$, of $P(m,n)$ coincides with the 
 distribution (derived in [20]) 
for the results $x$, $y$ of  measurements $\hat 
X_{\theta}^A$, $\hat X_{\phi}^B$ respectively, where  
$m\rightarrow \alpha x$, $n\rightarrow \beta y$. In the 
macroscopic limit the probability of outcome $0$ becomes $
\int_{-\delta_{0}}^{\delta_{0}}
dx\int_{-\infty}^{\infty}¥P_{a}¥(x,y)dy=\epsilon$ where 
$N_{0}=\alpha\delta_{0}$.
We can make this probability less than or equal to a
 pre-specified 
arbitrarily small value $\epsilon$, by determining $\delta_{0}$
from the asymptotic form $P_{a}¥(x,y)$, 
and choosing $\alpha\geq N_{0}/\delta_{0}¥$.

There is analogy with the Schr\"odinger gedanken experiment:  
a coupling of the entangled microscopic state $|\psi\rangle$ 
to ${\hat a_{2}}, \hat b_{2}$   
  generates the macroscopic 
 entangled state (4) (the cats). Subsequent 
 photon number measurements on these cats reveal a   
 macroscopic entanglement through the violation of the macroscopic Bell 
 inequalities (1). This macroscopic entanglement evident in $m,n$ 
   directly reflects the original continuous variable entanglement, 
 evident in $x,y$, of $|\psi\rangle$$^{\cite{10}}$. 
 
With this insight we predict that  
any state $|\psi\rangle$ demonstrating a failure of Bell's premise of 
local realism, for such continuous variable (amplitude or position and 
momentum) measurements, 
 will 
violate the macroscopic realism-locality premise we define here. 
A number of such entangled states have been recently 
predicted$^{\cite{20,21,22}}$ and are of increasing 
interest because of potential applications to the field of quantum 
information$^{\cite{19,23}}$.

    This experiment would then not seem impossible.   
Quantum states satisfying our criteria are two-mode equivalents to 
the coherent 
superposition states 
$\sim
(|\alpha_{0}\rangle \pm |-\alpha_{0}\rangle)/\sqrt{2}$,
the subject of much experimental interest, for small
 $\alpha_{0}¥$$^{\cite{3,4,5}}$. 
  As with balanced homodyne detection the photon number difference is  
measured by taking the 
difference of two currents generated from highly efficient photodiode 
detectors. The measurement apparatus is a simple modification of that 
proposed$^{\cite{24}}$ and used experimentally$^{\cite{19}}$ 
in the realization of the 
continuous variable Einstein-Podolsky-Rosen paradox$^{\cite{25}}$. 
We note that the quantum prediction 
could also apply to massive particles such as 
bosonic atoms.


\begin{references}
\bibitem{10} The measurements $S_{z}^{A}¥(\theta)$, $S_{z}^{A}¥(\theta')$
  are noncommuting: 
 $\Delta S_{x}^{A}\Delta S_{y}^{A}\geq \left| \left\langle S_z^A\right\rangle 
\right|/2 $ in the 
 Schwinger representation; there is no contradiction with an 
 uncertainty principle here, since $S_{z}^{A}¥=(\hat{a}^{\dagger}¥_{2}
\hat{a}_{2}-\hat{a}^{\dagger}_{1}\hat{a}_{1})/2$ becomes large in 
particle number; $S_{z}^{A}¥(\theta)=S_{x}^{A}cos\theta + S_{y}^{A}¥sin\theta$.
\bibitem{11} In the event of a violation of (1), it might be proposed 
that the macroscopic realism-locality premise fails because of an 
interaction of the local measurement procedure with the cat state. This 
possibility can be precluded because, as with the microscopic case,
 (1) is derivable assuming such interactions.
\bibitem{17}   While the 
    distinction $2N_{0}¥$ between our states is small relative to the 
    overall size $\alpha^{2}¥$ of the system, the absolute value of 
    this distinction can be as large as we specify.
\bibitem{18} Violations of a related premise, macroscopic local 
realism, are discussed in [9].
\end{references}
\end{document}